\documentclass{iaus}
\usepackage{graphicx}

\title[Carbon abundances in the Galactic thin and thick disks] 
{Carbon abundances in the \break Galactic thin and thick disks}

\author[T. Bensby \& S. Feltzing]   
{T. Bensby$^1$ \& S. Feltzing$^2$}%

\affiliation{$^1$Department of Astronomy, 830 Dennison Building, 
University of Michigan, \break Ann Arbor, MI 48109-1042, USA  
(email: tbensby@umich.edu)\\[\affilskip]
$^2$Lund Observatory, Box 43, 22100 Lund, Sweden (email: sofia@astro.lu.se)}

\pubyear{2005}
\volume{228}  
\pagerange{000--000}
\date{?? and in revised form ??}
\jname{From Li to U: Elemental Tracers of Early Cosmic Evolution}
\editors{V. Hill, P. Francois \& F. Primas, eds.}
\begin{document}

\maketitle

\begin{abstract}
Although carbon is, together with oxygen and nitrogen, one of the most
important elements in the study of galactic chemical evolution its
production sites are still poorly known and have been much debated 
(see e.g. Gustafsson et al. 1999; Chiappini et al. 2003).
To trace the origin and evolution of carbon we have
determined carbon abundances from the forbidden [C\,{\sc i}] line at
8727\,{\AA} and made comparisons to oxygen abundances from the
forbidden [O\,{\sc i}] line at 6300 {\AA} in a
sample of 51 nearby F and G dwarf stars. These data and the
fact that the forbidden [C\,{\sc i}] and [O\,{\sc i}] lines are very
robust abundance indicators (they are essentially insensitive to
deviations from LTE and uncertainties in the stellar parameters, see,
e.g., Gustafsson et al. 1999; Asplund et al. 2005) enable us to very
accurately measure the C/O ratio as well as individual C and O
abundances.  Our first results indicate that the time-scale for the
{\it main} source that contribute to the carbon enrichment of the
interstellar medium operate on the same time-scale as those that
contribute to the iron enrichment (and can possibly be AGB stars...)

\end{abstract}



\section{Observations, data reduction, and abundance analysis}
Spectra for 16 thick disk stars and 35 thin disk stars (based on
kinematical selection criteria, see Bensby et al. 2003, 2005) were obtained
during six nights in September 2004 with the CES spectrograph on the
ESO-3.6m telescope on La Silla. The spectra have a very high
resolution of $R\sim 220\,000$ and high signal-to-noise ratios of
$S/N\gtrsim300$.

Carbon abundances were determined by comparing synthetic spectra
(based on 1-D, plane parallel, LTE stellar atmosphere models that were
calculated with the Uppsala MARCS code, Gustafsson 1975; Edvardsson
1993) to the observed spectra. The method used for the analysis is the
same as that used for the forbidden oxygen line at 6300\,{\AA} (Bensby
et al. 2004).  Stellar parameters were taken from
Bensby et al.~(2003, 2005).  When synthesising the [C\,{\sc i}] line a
blending Fe\,{\sc i} line was taken into account (e.g., Lambert \& Swings 1967).  
Our analysis
of the solar spectrum gives a carbon abundance in the Sun of
$\log \epsilon (\rm C) = 8.41$, in good agreement with Allende Prieto
et al.~(2002) and Asplund et al.~(2005) who determined a solar
carbon abundance of 8.39 using 3D models.

\section{Results and discussion}

Our first results regarding carbon abundances in the Galactic thin
and thick disks are (Fig.~1): {\bf (1)} While the [O/Fe] vs [Fe/H] trends
separate well for the thin and thick disk samples, they are merged for
[C/Fe] vs [Fe/H] and are mainly flat; {\bf (2)} The two kinematic
samples separate well (as expected) in the [C/O] vs [O/H] plot; {\bf
(3)} The thin disk sample in the same plot have a nearly constant
[C/O] ratio; {\bf (4)} The thick disk sample first shows a flat trend
of [C/O] vs [O/H] for $\rm [O/H] < 0$ where C is depleted relative to
O such that $\rm [C/O]= -0.3$ to $-0.4$\,dex but at $\rm [O/H]=0$ the
trend rises (sharply) to solar values.

A preliminary interpretation of these results is as follows: {\bf (I)}
In the thin disk the sources for O (SN\,II) and C (debated) are tuned
such that a flat trend is created; {\bf (II)} In the thick disk the
processes are finely tuned up and until $\rm [O/H]=0$. At $\rm [O/H] =
0$ a new source of C gets active (or a source of O drops out); {\bf
(III)} Studying the [O/Fe] vs [Fe/H] trend we find that at $\rm
[Fe/H]\simeq-0.45$ SN\,Ia sets in (hence the appearance of the "knee"
in the abundance plot). For [C/Fe] this does not happen. This means
that the time-scale for the population that contributes C operates on
the same time-scale as SN\,Ia; {\bf (IV)} These results could be
exploited to derive an independent measure of the SN\,Ia lifetimes
(given an understanding of the sources of C).


%
\begin{figure}
\resizebox{\hsize}{!}{
 \includegraphics{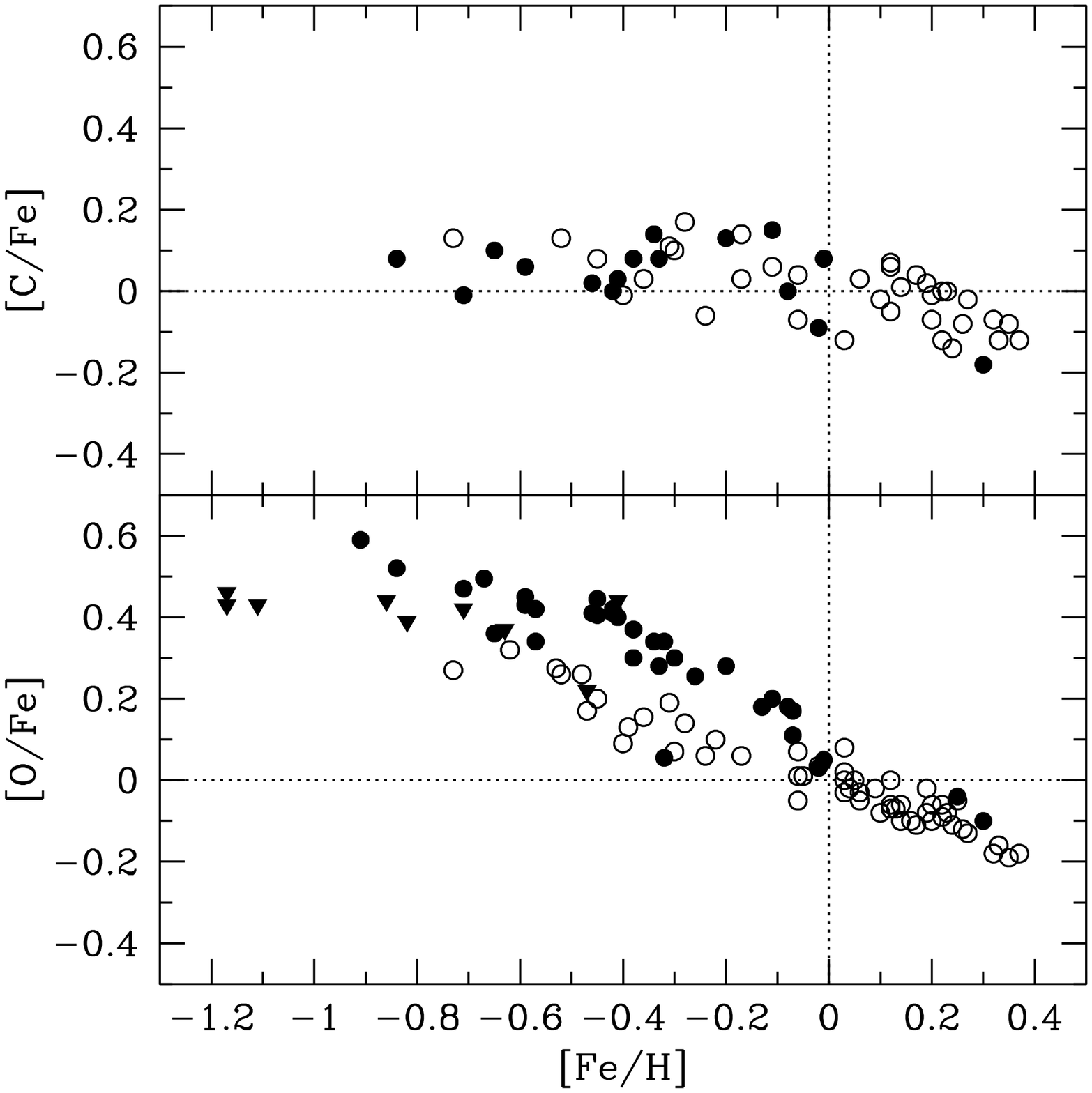}
 \includegraphics{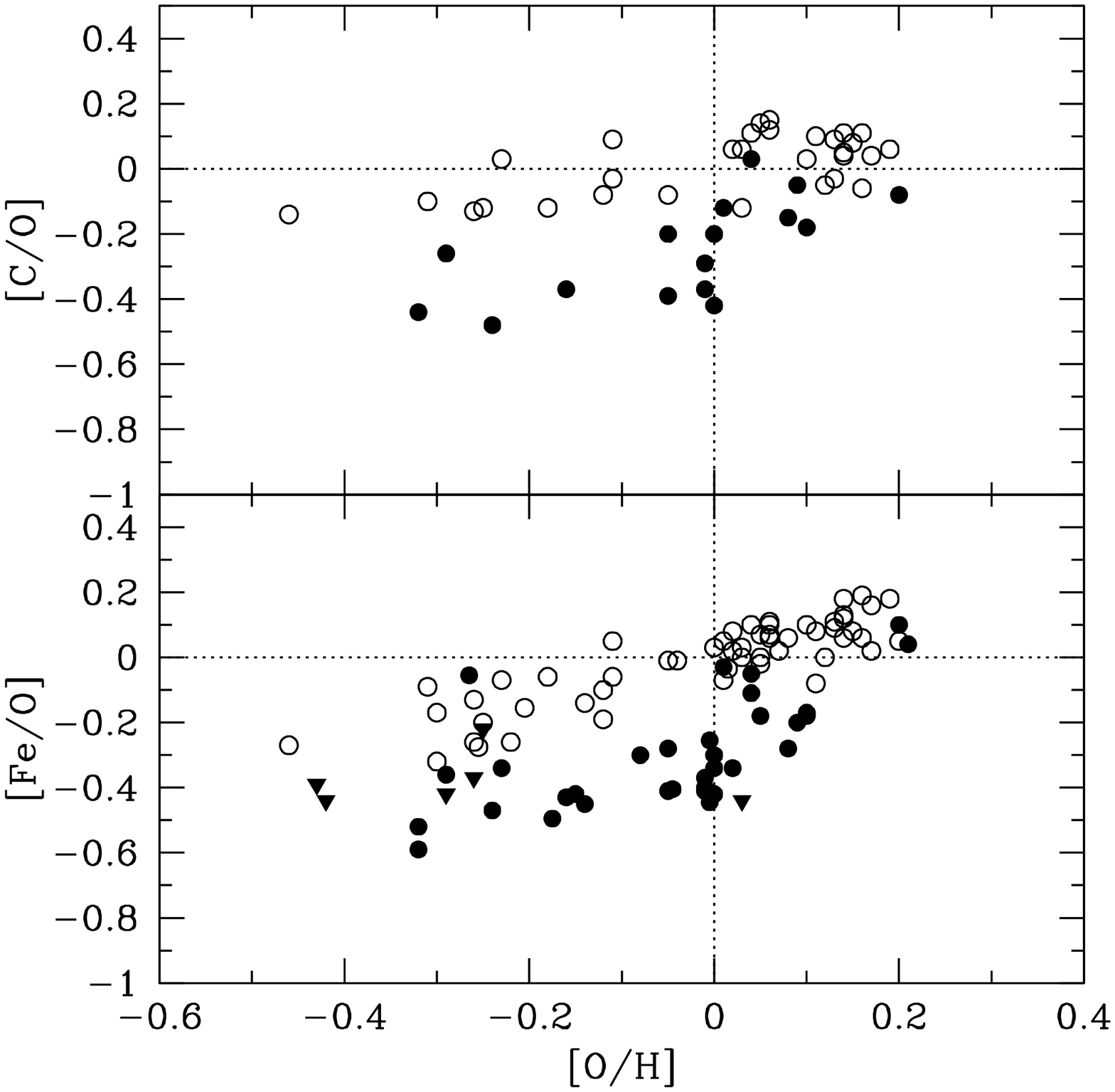}}
  \caption{
Our carbon trends relative to iron and relative to oxygen. For
comparison we also show the [O/Fe] vs [Fe/H] and [Fe/O] vs [O/H]
trends (Bensby et al. 2004, 2005). Thin and thick disk stars are
marked by open and filled symbols, respectively.
Filled triangles in the oxygen plots denote thick disks stars from 
Nissen et al. (2002). 
    }
\label{fig:wave}
\end{figure}
\begin{acknowledgments}
Thomas Bensby acknowledges support from the National Science Foundation, grant AST-0448900.
Sofia Feltzing is a Royal Swedish Academy Research Fellow supported by a grant 
from the Knut and Alice Wallenberg Foundation. 
\end{acknowledgments}


\begin{thebibliography}{}

\bibitem[2002]{allendeprieto}
Allende Prieto C., Lambert D.L., \& Asplund M., 2002, ApJ, 573, L137
\bibitem[2005]{asplund}
Asplund M., Grevesse N., Sauval A.J., Allende Prieto C., \& Blomme R., 2005, A\&A, 417, 751 
\bibitem[2003]{bensby03}
Bensby, T., Feltzing, S., \& Lundstr\"om, I., 2003, A\&A, 410, 527 
\bibitem[2004]{bensby04}
Bensby, T., Feltzing, S., \& Lundstr\"om, I., 2004, A\&A,  415, 155
\bibitem[2005]{bensby05}
Bensby, T., Feltzing, S., Lundstr\"om, I., \& Ilyin, I., 2005, A\&A, 433, 185 
\bibitem[2003]{chiappini}
Chiappini, C., Matteucci, F., \& Romano, D., 2003, MNRAS, 339, 63 
\bibitem[1993]{edvardsson}
Edvardsson, B., Andersen, J., Gustafsson, B. et al., 1993, A\&A, 275, 101
\bibitem[1975]{gustafsson1975}
Gustafsson, B., Bell, R.A., Eriksson, K., \& Nordlund, {\AA}., 1975, A\&A, 42, 407
\bibitem[1999]{gustafsson}
Gustafsson, B., Karlsson, T., Olsson, E., Edvardsson, B., \& Ryde, F., 1999, A\&A, 342, 426 
\bibitem[1967]{lambert} 
Lambert, D.L., \& Swings, J.P., 1967, Solar Physics, 2. 34
\bibitem[2002]{nissen}
Nissen, P.E., Primas, F., Asplund, M., \& Lambert, D.L., 2002, A\&A, 390, 235 


\end{thebibliography}
\end{document}